\documentclass[a4paper,11pt]{article}
\usepackage{pos}
\usepackage{graphicx}
\usepackage{wrapfig}
\usepackage[svgnames]{xcolor}
\usepackage[makeroom]{cancel}
\usepackage[normalem]{ulem}

\usepackage{booktabs}

\definecolor{blue-violet}{rgb}{0.54, 0.17, 0.89}
\definecolor{PineGreen}{cmyk}{0.92, 0, 0.59, 0.25}
\definecolor{OliveGreen}{cmyk}{0.64, 0, 0.95, 0.40}
\definecolor{RawSienna}{cmyk}{0, 0.72, 1, 0.45}
\definecolor{Gray}{cmyk}{0, 0, 0, 0.50}
\definecolor{MidnightBlue}{cmyk}{0.98, 0.13, 0, 0.43}
\definecolor{Orange}{cmyk}{0, 0.61, 0.87, 0}
\definecolor{LimeGreen}{cmyk}{0.50, 0, 1, 0}
\definecolor{Green}{cmyk}{1, 0, 1, 0}
\definecolor{brightube}{rgb}{0.82, 0.62, 0.91}

\title{Soft charges and zero modes at null boundaries}

\author[a]{Dusan Đorđević}
\author*[b,c]{Olivera Miskovic}
\author[b]{Antonia Montecinos}
\author[d]{Tatjana Vuka\v sinac\bigskip}

\affiliation[a]{Faculty of Physics, University of Belgrade,\\ Studentski Trg 12-16, 11000 Belgrade, Serbia}

\affiliation[b]{Instituto de F\'\i sica, Pontificia Universidad Cat\'olica de Valpara\'\i so,\\
Avenida Universidad 330, Curauma, Valpara\'{\i}so, Chile}

\affiliation[c]{Department of Applied Science and Technology, Politecnico di Torino,\\
Corso Duca degli Abruzzi 24, 10129 Torino, Italy}

\affiliation[d]{\it Facultad de Ingenier\'ia Civil, Universidad Michoacana de San Nicol\'as de Hidalgo, \\ Morelia, Michoac\'an 58000, Mexico \bigskip}

\emailAdd{dusan.djordjevic@ff.bg.ac.rs}
\emailAdd{olivera.miskovic@pucv.cl}
\emailAdd{antonia.montecinos@pucv.cl}
\emailAdd{tatjana.vukasinac@umich.mx}

\abstract{Asymptotic symmetries at null boundaries provide a powerful window into the soft sector of field theories, revealing conserved charges and infinite-dimensional symmetry algebras beyond the standard bulk dynamics. While these structures are usually associated with residual gauge transformations, null boundaries also exhibit a characteristic global zero-mode ambiguity, whose physical consequences are less explored. We discuss how this zero mode modifies the canonical structure
of field theories at null boundaries. In particular, we show that it gives rise to a new quasilocal edge observable, with consequences to the charge algebra. It can arise on any kind of null boundaries, both at the null infinity and the finite distance such as a black hole horizon.}

\FullConference{Foundations of General-Relativistic Gauge Field Theory (FGRGFT2026)\\
17-19 March, 2026\\
Politecnico di Torino, Italy\\}

 \tableofcontents

\begin{document}
\maketitle

\section{Null boundaries and global zero modes}

\begin{wrapfigure}[19]{l}{0.35\textwidth} \vspace{0pt} \centering \includegraphics[width=0.34\textwidth]{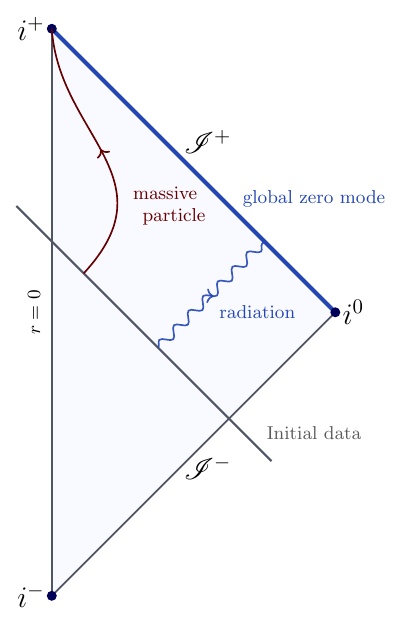} \caption{\footnotesize\itshape Global zero mode at the light front.} \label{fig:global-zero-mode} \vspace{-6pt} \end{wrapfigure}

Null boundaries play a special role in field theories and gravity. Unlike
spacelike surfaces, they are characteristic surfaces: part of the dynamics
is already encoded in the intrinsic data on the surface itself. This makes
null foliations particularly useful for studying radiation, soft modes and
the associated boundary observables.

A characteristic feature of null boundaries is the appearance of global zero modes \cite{Dirac:1949cp,Maskawa:1975hx,Heinzl:1998kz}. They are not propagating degrees of freedom, but they can
change the state of a field on the null surface without changing its energy.
In this sense, they are Goldstone-like. One therefore expects them to be related to soft charges and to shift symmetries intrinsic to the boundary \cite{Strominger:2017zoo}.
In the simple Minkowski example, the geometric meaning of this zero mode is
illustrated in Fig.~\ref{fig:global-zero-mode}. Namely, the zero mode is naturally tied to the light front, in a direction parallel to the initial-data surface. It therefore appears as a boundary shift of the field, rather than as a bulk gauge transformation
\cite{Alexandrov:2014rta}. 

The standard Dirac--Bergmann formalism is designed to identify gauge
symmetries generated by first-class constraints. In turn, the symmetries
considered here arise from residual zero modes of the symplectic matrix of
constraints after a null slicing has been chosen. They are not arbitrary functions in the bulk and therefore do not generate usual gauge transformations. Rather, they define residual transformations associated
with null surfaces.

The purpose of this note is to present a simple adaptation of the Dirac--Bergmann construction to null boundaries, thereby generalizing the results of Refs.~\cite{Gonzalez:2023yrz,Gonzalez:2024rho,DjMMV}. The idea is to study the
Hamiltonian evolution directly in a null foliation and to keep track of the non-uniqueness produced by residual zero modes of the constraint matrix related to the existence of centrally extended Kac-Moody algebras \cite{Floreanini:1987as,Salomonson:1988mk,Sazdovic:1995np,Miskovic:2001dw}.
When the corresponding generators are improved by Regge--Teitelboim boundary terms, the residual transformations give rise to soft charges. Depending on the boundary conditions and on the domain where the residual transformations
can be consistently extended, these charges can be interpreted either as global charges or as quasilocal charges associated with a chosen null patch.

\section{Canonical method for null surfaces}

The Dirac--Bergmann formalism for singular systems can be extended to
account for asymptotic symmetries that originate from residual zero modes of
the symplectic matrix of constraints, that are not arbitrary functions of all coordinates. The general form of this mechanism was
introduced in Ref.~\cite{Gonzalez:2024rho}. In what follows, we adapt this
construction to the case of asymptotic symmetries.

Let us assume that the relevant constraints are denoted by $\chi
_{\alpha}(u, x)\approx 0$ , where $x=[x^{i}]$, $i=1,\ldots ,d$, are coordinates on the
null hypersurface $\Xi $ defined by $u=\mathrm{const}$, and $\alpha =1,\ldots, k$. The symbol $\approx $ denotes weak equality, namely
equality on the constraint surface.  
The Poisson brackets
of these constraints define the operator 
\begin{equation}
\{\chi _{\alpha}(u,x),\chi _{\beta}(u,x')\}=\Omega _{\alpha\beta}(u;x,x')\,,
\end{equation}
which is antisymmetric under simultaneous exchange $(\alpha,x)\leftrightarrow
(\beta,x')$. Since $u$ is the evolution parameter in the Hamiltonian formalism, we shall suppress writing its explicit dependence.

We consider the case in which
the constraints $\chi_\alpha$ are second class at fixed point $x^i$ \cite{Nagarajan:1985xn,Goldberg:1991pb,Majumdar:2022fut}. Equivalently, the matrix $\Omega_{\alpha\beta}(x,x')$ is invertible, and its inverse $\Omega^{\alpha\beta}$ satisfies
$\Omega_{\alpha\gamma}\Omega^{\gamma\beta}= \delta_\alpha^\beta$ at fixed $x,x'$. This
does not, however, exclude the existence of residual zero modes that appear
when the kernel structure of $\Omega_{\alpha\beta}(x,x')$ is analyzed and its integral operator nature taken into account. These zero modes are not arbitrary functions on the whole hypersurface. If the boundary is located at  fixed $r=r_0$, where $x^{i}=(r,\varphi^A)$, they depend only on the boundary coordinates $\varphi^A$, and not on the radial coordinate $r$. Thus, they do not generate ordinary gauge symmetries in the bulk, but residual $r$-independent symmetries intrinsic to the boundary. As a consequence, the inverse matrix seen as an integral operator $\Omega^{\alpha\beta}(x,x')$ is not unique.

More precisely, we consider residual zero modes of the form 
\begin{equation}
v^\alpha(x)=\int\limits_{\partial \Xi }\mathrm{d}^{d-1}\varphi'\,P^\alpha_{\ I}(x,\varphi')\,V^{I}(\varphi')\,,  \label{factor}
\end{equation}
where $\partial\Xi$ corresponds to a boundary located at $r=\mathrm{const}$. Here $V^I(\varphi)$, $I=1,\ldots,K$ $(K \leq k)$, are arbitrary functions on $\partial\Xi$ and $P^\alpha_{\ I}(x,\varphi')$ is a fixed operator selecting the residual zero-mode
sector. Since the zero mode equation has the form
\begin{equation}
\int\limits_{\Xi }\mathrm{d}^{d}x'\,\Omega _{\alpha\beta}(x,x')v^\beta(x')=0\,,  \label{z.modes}
\end{equation}
valid for any $V^I(\varphi)$, the factorization \eqref{factor} is possible only if
the operators $\Omega _{\alpha\beta}$ and $P^\alpha_{\ I}$ satisfy the identity
\begin{equation}
\int\limits_{\Xi }\mathrm{d}^{d}x''\,\Omega _{\alpha\beta}(x,x'')P^{\beta}_{\ I}(x'',\varphi') \equiv 0\,.  \label{P.property}
\end{equation}
Hence, the operator $P^\alpha_{\ I}(x,\varphi')$ may be regarded as the map that embeds the residual zero-mode sector, labeled by $(I,\varphi^A)$, into the original space labeled
by the pair $(\alpha,x^i)$. Equivalently, it projects the full symplectic matrix
onto the residual sector, $\Omega_{\alpha\beta}(x,x') \to \Omega_{IJ}(\varphi,\varphi')$, where it vanishes: 
\begin{equation}
\Omega_{IJ}(\varphi,\varphi')
\equiv
\int\limits_{\Xi}\mathrm{d}^d x''
\int\limits_{\Xi}\mathrm{d}^d x'''\,
P^\alpha_{\ I}(x'',\varphi)P^\beta_{\ J}(x''',\varphi')\, \Omega_{\alpha\beta}(x'',x''')=0 .
\end{equation}
Therefore, the constraints $\chi _\alpha(x)$ remain second class in the discrete indices $\alpha$, while the non-uniqueness, associated with the
residual kernel of $\Omega_{\alpha\beta}$ due to $\Omega_{IJ}=0$, is the origin of the asymptotic symmetry.

The properties discussed so far are purely kinematical. To introduce the dynamics, let the Hamiltonian be 
\begin{equation}
H=\int\limits_{\Xi }\mathrm{d}^{d}x\,\left( \mathcal{H}_{0}+\lambda ^\alpha\chi
_\alpha\right) \,,
\end{equation}
where $\mathcal{H}_{0}$ is the multiplier-independent part and $\lambda^{\alpha}(x)$ are Hamiltonian multipliers. For simplicity, we suppose that there are no additional constraints. Preservation of the constraints under the
evolution generated by $H$ gives 
\begin{equation}
\dot{\chi}_\alpha(x)=\int\limits_{\Xi }\mathrm{d}^{d}x'\,\{\chi
_\alpha(x),\mathcal{H}_{0}(x')\}+\int\limits_{\Xi }\mathrm{d}^{d}x'\,\Omega _{\alpha\beta}(x,x')\lambda ^{\beta}(x')\approx
0\,.
\end{equation}%
Equivalently, the multipliers satisfy the equation 
\begin{equation}
\int\limits_{\Xi }\mathrm{d}^{d}x'\,\Omega _{\alpha\beta}(x,x')\lambda ^{\beta}(x')=J_{\alpha}(x)\,,  \label{eq.lambda}
\end{equation}
with the source corresponding to a non-homogeneous part of the equation, 
\begin{equation}
J_\alpha(x)=\int\limits_{\Xi }\mathrm{d}^{d}x'\,\{\mathcal{H}_{0}(x'),\chi_\alpha(x)\}\,.
\end{equation}
Because of the residual zero modes, the solution for the multipliers is not
unique. Using \eqref{P.property}, its general form is 
\begin{equation}
\lambda ^\alpha(x)=\bar{\lambda}^\alpha(x;J)+\int\limits_{\partial \Xi }\mathrm{d}^{d-1}\varphi'\,P^\alpha_{\ I}(x,\varphi')\Lambda ^I(\varphi')\,,
\end{equation}
where $\bar{\lambda}^{\alpha}(x;J)$ is a particular solution of \eqref{eq.lambda}, while $\Lambda ^I(\varphi)$ are arbitrary boundary functions. The arbitrary part of $\lambda ^{\alpha}$ therefore does not represent a gauge ambiguity in the whole spacetime, because the corresponding parameters are not arbitrary functions in the bulk. Rather, it defines residual
transformations associated with the surfaces $r=\mathrm{const}$.

Since the Regge--Teitelboim charges are surface integrals \cite{Regge:1974zd}, these residual transformations become boundary symmetries once appropriate boundary
conditions are imposed. In this sense, the residual symmetry is an intrinsic
symmetry of the boundary at $r=r_0$, and the corresponding charge is
quasilocal. If the boundary conditions allow the residual transformation to
be extended consistently to the whole region under consideration, the same
charge may be interpreted as a global charge.

The corresponding residual
generator can be written as 
\begin{equation}
G[\varepsilon ]=\int\limits_{\Xi }\mathrm{d}^{d}x\,\chi _\alpha(x)\eta
^\alpha(x)\,,\qquad \eta ^\alpha(x)=\int\limits_{\partial \Xi }\mathrm{d}^{d-1}\varphi'\,P^\alpha_{\ I}(x,\varphi')\varepsilon ^I(\varphi')\,.
\label{G}
\end{equation}
The parameters $\eta ^\alpha(x)$ are not arbitrary gauge parameters, but they
are restricted to lie in the residual zero-mode sector, such that the independent parameters are the boundary functions $\varepsilon ^I(\varphi)$. Using the property \eqref{z.modes} satisfied by the parameters $\eta^\alpha(x)$ due to
\eqref{P.property}, one finds that the bulk contribution to the Poisson bracket of two such generators vanishes, 
\begin{eqnarray}
\{G[\varepsilon _{1}],G[\varepsilon _{2}]\} &\approx &\int\limits_{\Xi } \mathrm{d}^{d}x\int\limits_{\Xi } 
\mathrm{d}^{d}x'\,\Omega
_{\alpha\beta}(x,x')\,\eta
^{\alpha}(x)\eta ^{\beta}(x') = 0\,,
\end{eqnarray}
up to possible boundary terms. The weak equality arises from the possible dependence of $P^\alpha{}_{I}$ on the fields. We also assume that $\varepsilon^I$ are field-independent. Thus, $G[\varepsilon ]$ indeed generates a residual asymptotic symmetry rather than an ordinary bulk gauge symmetry. 

The rest of the procedure is the same as in the standard Dirac--Bergmann
formalism. The functional $G[\varepsilon ]$ need not be differentiable
because boundary terms have been omitted in the computation of Poisson
brackets. Following the Regge--Teitelboim prescription \cite{Regge:1974zd},
one improves it by adding a boundary charge, 
\begin{equation}
G_{Q}[\varepsilon ]=G[\varepsilon ]+Q[\varepsilon ]\,,
\end{equation}%
so that $G_{Q}[\varepsilon ]$ has well-defined functional derivatives. The
algebra of improved generators may then acquire a boundary contribution, 
\begin{equation}
\{G_{Q}[\varepsilon _{1}],G_{Q}[\varepsilon _{2}]\} = G_{Q}[[\varepsilon
_{1},\varepsilon _{2}]]+ C[\varepsilon _{1},\varepsilon _{2}]\,,
\label{GG=C}
\end{equation}
where the parameter $[\varepsilon _{1},\varepsilon
_{2}]$ is defined by the structure constants of the algebra and $C[\varepsilon_1,\varepsilon_2]$ is a possible central charge. We will show in the next section that, for the class of examples considered below, $G_Q[[\varepsilon _{1},\varepsilon
_{2}]]=0$ and the possible central term is given by the pullback of the symplectic form to the residual zero-mode sector, and that this pullback vanishes. Consequently, the residual shift charges form an Abelian algebra without central extension.

\section{Application}

All examples considered so far in Refs.~\cite{Gonzalez:2023yrz,Gonzalez:2024rho,DjMMV}, for $d=2$ and $d=3$, have some
common features. Let us summarize them in $(d+1)$-dimensional Minkowski
spacetime, with the asymptotic boundary located at $r_{0}=\infty $. We use
retarded or advanced Bondi coordinates, in which the Minkowski line element
takes the form 
\begin{equation}
\mathrm{d}s^{2}=-\mathrm{d}u^{2}-2\epsilon \,\mathrm{d}u\,\mathrm{d}
r+r^{2}\gamma _{AB}(\varphi)\,\mathrm{d}\varphi^{A}\mathrm{d}\varphi^{B}\,,
\end{equation}%
where $\epsilon =+1$ for retarded time and $\epsilon =-1$ for advanced time.
The coordinates $\varphi^{A}$ are local coordinates on the unit $(d-1)$-sphere $\mathbb{S}^{d-1}$.

Let the fields be denoted collectively by $\Psi^\Lambda\in \{\phi,A_{\mu },A_{\mu }^{a}\}$, with canonical momenta $\Pi _{\Lambda }$. We also
denote by $\psi ^{\alpha }\in \{\phi ,A_{A},A_{A}^{a}\}$ the particular
subset of fields affected by the global zero modes, with canonical momenta $\pi_\alpha$. The primary constraints
characteristic of the null foliation have the form 
\begin{equation}
\chi _{\alpha }=\pi _{\alpha }+f_{\alpha }(\Psi )\approx 0\,,
\label{primary}
\end{equation}
where $f_\alpha$ depends on $\Psi,\partial_i\Psi,\ldots$.
 The Poisson brackets of the constraints define the operator 
\begin{equation}
\{\chi _{\alpha }(x),\chi _{\beta }(x')\}=\Omega _{\alpha \beta
}(x,x')\equiv \frac{\delta f_{\alpha }(x)}{\delta \psi ^{\beta }(x')}-\frac{\delta f_{\beta }(x')}{
\delta \psi ^{\alpha }(x)}\,.
\end{equation}
In all the considered cases, $\Omega _{\alpha \beta }(x,x')$ is an operator linear in derivatives. It can therefore be written in the
form 
\begin{equation}
\Omega _{\alpha \beta }(x,x')=\hat{\mathcal{L}}_{\alpha \beta
}(x)\,\delta ^{(d)}(x-x')\,,
\label{Omega}
\end{equation}
where $\hat{\mathcal{L}}_{\alpha \beta }(x)$ is a differential operator
linear in derivatives along $\Xi $.

The zero modes of $\Omega _{\alpha \beta }$ are functions $v^{\alpha }(x)$
satisfying \eqref{z.modes}. For a local operator of the form above, this condition
reduces to the linear differential equation
\begin{equation}
\hat{\mathcal{L}}_{\alpha \beta }(x)v^{\beta }(x)=0\,.
\end{equation}%
In the examples considered here, the corresponding zero-mode map is also
local in the angular variables. Namely, it has the form 
\begin{equation}
P_{\ I}^{\alpha }(x,\varphi')=U_{\ I}^{\alpha }(r,\varphi)\,\delta
^{(d-1)}(\varphi-\varphi')=U_{\ I}^{\alpha }(x)\,\delta
^{(d-1)}(\varphi-\varphi')\,.
\label{P}
\end{equation}
The property \eqref{P.property} then implies 
\begin{equation}
\hat{\mathcal{L}}_{\alpha \beta }U^{\beta }_{\ I}=0\,,
\end{equation}
where $U^{\alpha}_{\ I}$ does not depend on $(\psi^\alpha, \pi_\beta)$, but it can depend on other fields.\footnote{In Yang-Mills theory, for example, $U$ depends on the radial gauge field $A^a_r$, while the components affected by the zero modes are the angular ones, $A_A^a$.} Therefore, the zero modes are parametrized as 
\begin{equation}
v^{\alpha }(x)=U_{\ I}^{\alpha }(x)\,V^{I}(\varphi)\,,
\end{equation}
where $V^{I}(\varphi)$ is an arbitrary function on the angular section of the null boundary. As we mentioned previously, all quantities, including zero modes, can also depend on $u$, which we do not write explicitly for simplicity of notation. The matrix $U^{\alpha }_{\ I}(x)$ determines the radial behavior of the zero modes, or equivalently of the residual symmetry parameters, near
the boundary $r=r_{0}$.

The assumptions used in \eqref{primary}, \eqref{Omega} and \eqref{P} are not restrictive. They only encode the standard local structure of the field theory examples under consideration. The form \eqref{primary} follows from the fact that, in a null foliation, the second order Lagrangians become linear in the velocities of the boundary fields $\psi^\alpha$ \cite{Alexandrov:2014rta,Steinhardt:1979it} . The local form \eqref{Omega} follows from locality of the
action, while \eqref{P} expresses the fact that, in these examples, the residual zero modes are local in the angular variables and have a radial dependence fixed by the zero-mode equation.

This simple local structure is a special feature of the scalar, Maxwell, scalar-Maxwell, Maxwell-Pontryagin, axion-Maxwell and Yang--Mills examples considered here.

The improved generator \eqref{G} of the residual asymptotic symmetry can
then be written as 
\begin{equation}
G_{Q}[\varepsilon] = \int\limits_{\Xi}\mathrm{d}^{d}x\,
\chi_{\alpha}(x)\eta^{\alpha}(x) + Q[\varepsilon]\,, \qquad \eta^{\alpha}(x)
= U^{\alpha}_{\ I}(x)\,\varepsilon^{I}(\varphi)\,.  \label{GQlocal}
\end{equation}
Here $\varepsilon^{I}(\varphi)$ is the residual symmetry parameter at the null
boundary, while its radial dependence is fixed by the zero-mode condition. 

In gauge theories there are, of course, also the usual gauge generators, such
as the Gauss constraint in Maxwell and Yang--Mills theory. In what follows we
do not discuss the full asymptotic symmetry algebra. We isolate only the
subalgebra generated by the residual zero modes of the null constraint matrix,
which is the sector responsible for the soft charges considered here.

The transformation of the field component associated with
the zero mode is of shift type in $\eta$, 
\begin{equation}
\delta_{\varepsilon}\psi^{\alpha}(x) =\{ \psi^{\alpha}(x), G_{Q}[\varepsilon] \}  = \eta^{\alpha}(x)= U^{\alpha}_{\ I}(x)\,\varepsilon^{I}(\varphi)\,.
\label{var.eta}
\end{equation}
The parameter $\eta^{\alpha}(x)$ is restricted by the condition that it be a zero mode of the operator $\Omega_{\alpha\beta}$. Let us emphasize again that this transformation is not, in general, an
arbitrary bulk gauge transformation. Rather, it is the residual
transformation selected by the zero-mode condition at the null boundary.

\begin{table}[t]
\centering
\renewcommand{\arraystretch}{1.45}
\setlength{\tabcolsep}{6pt}
\small
\begin{tabular}{@{}lccc p{3.0cm} p{3.0cm} p{2.4cm}@{}}
\hline
Theory
&
$D$
&
$\Psi^\Lambda$
&
$\psi^\alpha$
&
$f_\alpha$
&
$\hat{\mathcal L}_{\alpha\beta}$
&
$U^\alpha{}_{I}$
\\
\hline

Scalar
&
$4$
&
$\phi$
&
$\phi$
&
$-\epsilon r^{2}\sqrt{\gamma}\,\partial_{r}\phi$
&
$-2\epsilon\sqrt{\gamma}\,\partial_{r}(r\,\cdot)$
&
$r^{-1}$
\\[2mm]

Maxwell--Pontryagin
&
$4$
&
$A_\mu$
&
$A_A$
&
$-\epsilon\sqrt{\gamma}\,\sigma^{AB}F_{rB}$
&
$-\dfrac{2\epsilon}{e^{2}}\sqrt{\gamma}\,
\gamma^{AB}\partial_{r}$
&
$\delta^{A}_{B}$
\\[3mm]

Maxwell
&
$3$
&
$A_\mu$
&
$A_\varphi$
&
$-\dfrac{\epsilon}{e^{2}}F_{r\varphi}$
&
$-\dfrac{2\epsilon}{e^2\sqrt r}\,
\partial_r(r^{-1/2}\,\cdot)$
&
$\sqrt r$
\\[3mm]

Yang--Mills
&
$4$
&
$A_\mu^{a}$
&
$A_A^{a}$
&
$-\dfrac{\epsilon}{g^{2}}\sqrt{\gamma}\,
\gamma^{AB}F_{rB}^{a}$
&
$-\dfrac{2\epsilon}{g^{2}}\sqrt{\gamma}\,
\gamma^{AB}(\mathcal D_{r})^{a}{}_{b}$
&
$U^{a}{}_{b}\delta^{A}_{B}$
\\

\hline
\end{tabular}
\caption{Primary constraints $\chi_\alpha=\pi_\alpha+f_\alpha$ and zero-mode operators $\hat{\mathcal L}_{\alpha\beta}$ in the examples
considered. Here $D=d+1$ denotes the spacetime dimension. The matrix $U^\alpha{}_{I}$ determines the radial dependence of the residual zero modes.}
\label{tab:zero-mode-operators}
\end{table}

In the examples at hand, the operators are given in Table~\ref{tab:zero-mode-operators}.
Here,
\begin{equation}
\sigma^{AB}
=
\frac{1}{e^{2}}\,\gamma^{AB}
-\frac{\epsilon\theta}{e^{2}\sqrt{\gamma}}\,\epsilon^{AB}\,,
\end{equation}
for the Maxwell--Pontryagin theory, up to the convention used for the normalization
of the Pontryagin coupling. With this definition, the $\theta$-dependent antisymmetric part of
$\sigma^{AB}$ does not contribute to the symmetric Poisson bracket operator
$\hat{\mathcal L}_{AB}$, which is why the latter is the same as in pure Maxwell theory.

For the Yang--Mills row, $U^{a}_{\ b}$ is defined by
\begin{equation}
(\mathcal D_r)^a_{\ c}U^c_{\ b}=0\,,
\qquad
\lim_{r\to\infty}U^a_{\ b}(r,\varphi)=\delta^a_b \, .
\end{equation}
Equivalently, in matrix notation,
\begin{equation}
U(r,\varphi)
=
\mathcal P\exp\left(
\int_{r}^{\infty}\mathrm{d}r'\,A_r(r',\varphi)
\right)\,, \qquad
A_r=A_r^aT_a \,,
\end{equation}
which is the radial Wilson line, if the radial covariant derivative acts as
$\mathcal D_r=\partial_r+[A_r,\,\cdot\,]$. Note that, in this case, the action of
the residual map $P^a_{\ b}(x,\varphi')$ on a boundary parameter $V_A^a(\varphi)$ defines the zero modes as
\begin{equation}
v^a_A(x)
=
\int\limits_{\partial\Xi}\mathrm{d}^{d-1}\varphi'\,
P^a_{\ b}(x,\varphi')\,V^b_A(\varphi')
=
\bigl[U(x)V_A(\varphi)U^{-1}(x)\bigr]^a \, .
\end{equation}
Thus, $P^a_{\ b}$ is  a field-dependent operator built from the radial Wilson line. 

The term $f_\alpha$ is the part of the constraint that is relevant for the definition of the conserved charges. In the considered examples, it is
linear in derivatives and can be written in the form
\begin{equation}
f_\alpha
= k_\alpha(\Psi) -
k_{\alpha\Lambda}^i(\Psi)\,\partial_i\Psi^\Lambda\,,
\end{equation}
where $k_{\alpha\Lambda}^i$ and $k_\alpha$ do not depend on
derivatives of the fields. Since the residual zero-mode operator
$U^\alpha_{\ I}$ does not contain derivatives, the non-differentiable part
of the variation of the generator comes only from the radial derivative term.
Thus, one finds
\begin{equation}
\delta G[\varepsilon]
=
\hbox{regular terms}
-
\int\limits_{\partial\Xi}\mathrm{d}^{d-1}\varphi\,
k_{\alpha\beta}^{r}\,
\eta^\alpha\,
\delta\psi^\beta ,
\end{equation}
where it was accounted that only the fields $\psi^\beta$ carry boundary degrees of freedom. The Regge--Teitelboim boundary term is therefore defined by
\begin{equation}
\delta Q[\varepsilon]
=
\int\limits_{\partial\Xi}\mathrm{d}^{d-1}\varphi\,
k_{\alpha\beta}^{r}\,
\eta^\alpha\,
\delta\psi^\beta ,
\qquad
\eta^\alpha
=
U^\alpha_{\ I}\,\varepsilon^I .
\end{equation}
This expression is not integrable in general because both $k_{\alpha\beta}^{r}$ and $U^\alpha_{\ I}$ may depend on
the fields. We shall impose boundary conditions such that their leading
boundary values are fixed,
\begin{equation}
k_{\alpha\beta}^{r}
\longrightarrow
\bar k_{\alpha\beta}^{r}\,,
\qquad
U^\alpha_{\ \ I}
\longrightarrow
\bar U^\alpha_{\ \ I}\,,
\qquad
r\to r_0 \,,
\label{b.background}
\end{equation}
with
\begin{equation}
\delta\bar k_{\alpha\beta}^{r}=0\,,
\qquad
\delta\bar U^\alpha_{\ \ I}=0\, .
\end{equation}
Under these assumptions, the quasilocal charge becomes integrable and takes the form
\begin{equation}
Q[\varepsilon]
=
\int\limits_{\partial\Xi}\mathrm{d}^{d-1}\varphi\,
\bar k_{\alpha\beta}^{r}\,
\bar U^\alpha_{\ I}\,
\varepsilon^I\,
\psi^\beta \, .
\label{Q}
\end{equation}

For the field theory examples considered here, the relevant
non-differentiable terms and the corresponding boundary data are summarized
in Table~\ref{tab:boundarydata}. In all cases $k_{\alpha\beta}^{r}$ is
field independent, so that $k_{\alpha\beta}^{r}=\bar k_{\alpha\beta}^{r}$
directly. The same is true for $U^\alpha_{\ I}$, except in Yang--Mills
theory. In that case, the residual zero-mode operator is built from the radial
Wilson line. Since $A_r^a=\mathcal O(r^{-2})$ \cite{Strominger:2013lka,He:2015zea}, the Wilson line approaches
the identity at the asymptotic boundary, and therefore
$\bar U^{aA}_{\ bB}=\delta^a_b\delta^A_B$.

\begin{table}[t]
\centering
\renewcommand{\arraystretch}{1.45}
\setlength{\tabcolsep}{6pt}
\small
\begin{tabular}{@{}l c p{3.0cm} p{3.0cm} p{2.4cm}@{}}
\hline
Theory
&
$\psi^\alpha$
&
$k_{\alpha\beta}^{r}\partial_r\psi^\beta$
&
$\bar k_{\alpha\beta}^{r}$
&
$\bar U^\alpha{}_{I}$
\\
\hline

Scalar
&
$\phi$
&
$\epsilon r^2\sqrt{\gamma}\,\partial_r\phi$
&
$\epsilon r^2\sqrt{\gamma}$
&
$r^{-1}$
\\[2mm]

Maxwell--Pontryagin
&
$A_A$
&
$\epsilon\sqrt{\gamma}\,\sigma^{AB}\partial_r A_B$
&
$\epsilon\sqrt{\gamma}\,\sigma^{AB}$
&
$\delta^A_{B}$
\\[3mm]

Maxwell
&
$A_\varphi$
&
$\dfrac{\epsilon}{e^2}\,\partial_r A_\varphi$
&
$\dfrac{\epsilon}{e^2}$
&
$\sqrt r$
\\[3mm]

Yang--Mills
&
$A_A^a$
&
$\dfrac{\epsilon}{g^2}\sqrt{\gamma}\,
\gamma^{AB}g_{ab}\,\partial_r A_B^b$
&
$\dfrac{\epsilon}{g^2}\sqrt{\gamma}\,
\gamma^{AB}g_{ab}$
&
$\delta^a_{b}\delta^A_{B}$
\\

\hline
\end{tabular}
\caption{Boundary data entering the Regge--Teitelboim charges. Here, $g_{ab}$ is the Cartan-Killing metric of the Lie algebra.}
\label{tab:boundarydata}
\end{table}

For the charge \eqref{Q}, the corresponding charge algebra in the reduced phase space is most directly obtained from
the canonical symplectic form $\omega =\int_{\Xi }\mathrm{d}^{d}x$ $\delta
\Pi _{\Lambda }\wedge \delta \Psi ^{\Lambda }$. The vector field tangent to
the reduced phase space has the form 
\begin{equation}
X_{\varepsilon }=\int\limits_{\Xi }\mathrm{d}^{d}x\,\left( \delta
_{\varepsilon }\psi ^{\alpha }\,\frac{\delta }{\delta \psi ^{\alpha }}%
+\delta _{\varepsilon }\pi _{\alpha }\,\frac{\delta }{\delta \pi _{\alpha }}%
\right) \,,
\end{equation}
where the  transformation of the field $\psi^\alpha$ is given by \eqref{var.eta}, while the
transformation of the momenta is not independent. Namely, on the reduced phase space the constraints vanish, and the vector field must be tangent to the
constraint surface, $\delta _{\varepsilon }\chi _{\alpha }(x)=0$. Hence 
\begin{equation}
\delta _{\varepsilon }\pi _{\alpha }(x)=\int\limits_{\Xi }\mathrm{d}
^{d}x'\,\frac{\delta f_{\alpha }(x)}{\delta \psi ^{\beta
}(x')}\,\eta ^{\beta }(x')\,.
\end{equation}
Then the charge is obtained from the exact form $i_{X_{\varepsilon }}\omega $
as 
\begin{equation}
i_{X_{\varepsilon }}\omega =-\delta Q[\varepsilon ]\,,
\end{equation}%
whereas the charge algebra is its double contraction, 
\begin{equation}
\{Q[\varepsilon _{1}],Q[\varepsilon _{2}]\}^{\ast }=-i_{X_{\varepsilon
_{1}}}i_{X_{\varepsilon _{2}}}\omega \,.
\end{equation}
By direct evaluation and using $\delta_\varepsilon U^{\alpha}_{\ I}=0$ since $U$ does not depend on the fields that transform non-trivially under the shift transformations, one finds that the above expression vanishes,
\begin{align}
-i_{X_{\varepsilon _{1}}}i_{X_{\varepsilon _{2}}}\omega & =-\int\limits_{\Xi
}\mathrm{d}^{d}x\,\left( \delta _{\varepsilon _{2}}\pi _{\alpha }\,\eta
_{1}^{\alpha }-\delta _{\varepsilon _{1}}\pi _{\alpha }\,\eta _{2}^{\alpha
}\right)   \notag \\
& =\int\limits_{\Xi }\mathrm{d}^{d}x\int\limits_{\Xi }\mathrm{d}%
^{d}x'\,\eta _{1}^{\alpha }(x)\left[ \frac{\delta f_{\alpha }(x)}{%
\delta \psi ^{\beta }(x')}-\frac{\delta f_{\beta }(x')}{%
\delta \psi ^{\alpha }(x)}\right] \eta _{2}^{\beta }(x')  \notag \\
& =\int\limits_{\Xi }\mathrm{d}^{d}x\int\limits_{\Xi }\mathrm{d}
^{d}x'\,\eta _{1}^{\alpha }(x)\Omega _{\alpha \beta }(x,x^{\prime
})\eta _{2}^{\beta }(x')=0\,,
\end{align}%
because the parameters $\eta ^{\alpha }$ are residual
zero modes of $\Omega _{\alpha \beta }$. Therefore 
\begin{equation}
\{Q[\varepsilon _{1}],Q[\varepsilon _{2}]\}^{\ast }=0\,.
\end{equation}
We conclude that the charge algebra is Abelian,  $Q[[\varepsilon_1,\varepsilon_2]]=0$, and the pullback of the symplectic
matrix to the residual zero-mode sector vanishes, $\Omega _{IJ}(\varphi
,\varphi ')=0$. Thus the residual shift charges form an Abelian algebra without central extension.  In Eq.~\eqref{GG=C}, it means that $G_Q[[\varepsilon_1,\varepsilon_2]]=0$ and $C[\varepsilon _{1},\varepsilon _{2}]=0$.

Let us finally comment on a case where the null boundary is located at finite distance, namely the horizon of the BTZ black hole \cite{Banados:1992wn,Carlip:1995qv}. The relevant difference with respect to Minkowski space is the location of the null boundary. In Minkowski space, the null boundary is located at infinity, whereas in the BTZ geometry the conformal boundary of anti-de Sitter space is timelike. Therefore, the natural null boundary for the present construction is the black hole horizon.

In both the Minkowski and BTZ cases, the Regge--Teitelboim surface term defines a quasilocal charge associated with the chosen null boundary \cite{DjMMV}. Since the residual parameters are arbitrary functions of the boundary coordinates, the corresponding shift symmetries form an infinite-dimensional Abelian algebra intrinsic to the null surface. This is in line with the appearance of infinite-dimensional symmetry algebras both at asymptotic null boundaries in gauge theories \cite{Strominger:2013lka,He:2015zea} and at black hole horizons \cite{Donnay:2016ejv}.

\section{Discussion}

We have described a simple extension of the Dirac--Bergmann construction adapted to null foliations. The main point is that null evolution can produce residual zero modes of the symplectic matrix of constraints. These zero modes are not arbitrary functions in the bulk and therefore do not
correspond to ordinary gauge transformations. Instead, they generate residual shift symmetries intrinsic to the null boundary. After the Regge--Teitelboim improvement, the corresponding generators reduce on the constraint surface to boundary charges. 

For each null patch, labeled by $\epsilon =\pm 1$, the construction gives a boundary charge $Q[\varepsilon ]$. Thus, before imposing matching
conditions, one has two copies of the charge, one associated with the future patch and one with the past patch. Matching conditions are needed to relate
these two copies at the endpoints of the patches and to formulate a conservation law on the completed null surface. In Minkowski spacetime this relation is implemented by the usual antipodal matching conditions imposed on the canonical fields and on the residual parameters \cite{Kapec:2015ena,Campiglia:2015qka,Campiglia:2017dpg,Strominger:2017zoo,Campiglia:2018see}, relating the data at $\mathscr I_{-}^{+}$ and $\mathscr I_{+}^{-}$. For a finite null boundary, such as a black hole horizon, there is no universal antipodal map. Instead, the physical quantities at the endpoint $r=r_{+}$, $u\rightarrow -\infty $ of the future patch are matched with the corresponding endpoint $r=r_{+}$, $%
u\rightarrow +\infty $ of the past patch, with the angular identification fixed by the horizon geometry. 

In a single patch, conservation is more properly expressed as a flux-balance equation. If $u=u_1$ and $u=u_2$ are two sections of the same null boundary, then the change of the charge between them takes the schematic form 
\begin{equation}
Q[\varepsilon ]\big|_{u_{2}}-Q[\varepsilon] \big|_{u_{1}}=\int\limits_{u_{1}}^{u_{2}}\mathrm{d}u\,\mathcal{F}[\varepsilon](u)\,,
\end{equation}
where $\mathcal{F}[\varepsilon](u)=\dot{Q}[\varepsilon]$ is the flux
associated with the residual symmetry. Its explicit form depends on the system and on the boundary conditions. In the black hole case, this equation describes the flux of the soft charge across the horizon. This suggests that the same zero-mode mechanism can provide a unified description of soft charges both at null infinity and at finite null boundaries.

\section*{Acknowledgments}

This work was partially funded by FONDECYT Regular Grants No.~1230492 and No.~1231779. The work of D.D. was supported by the Faculty of Physics, University of Belgrade, through grant No.~451-03-137/2025-03/200162 of the Ministry of Science, Technological Development and Innovations of the Republic of Serbia. T.V.~acknowledges financial support from CIC, Universidad Michoacana de San Nicol\'as de Hidalgo, Mexico. A.M.~acknowledges financial support from Pontificia Universidad Cat\'olica de Valpara\'iso through the PAIM scholarship.

\end{document}